\documentclass[12pt,a4paper]{article}

\newcommand{\sg}{\sqrt{2}g}
\newcommand{\et}{e^{-|t|}}
\newcommand{\ep}{\epsilon}

\begin{document}

\begin{flushright}
{ }
\end{flushright}
\vspace{1.8cm}

\begin{center}
 \textbf{\Large Analytic Bethe Ansatz Solutions for Highest States \\
 in the $su(1|1)$ and $su(2)$ Sectors }  
\end{center}
\vspace{1.6cm}
\begin{center}
 Shijong Ryang
\end{center}

\begin{center}
\textit{Department of Physics \\ Kyoto Prefectural University of Medicine
\\ Taishogun, Kyoto 603-8334 Japan}
\par
\texttt{ryang@koto.kpu-m.ac.jp}
\end{center}
\vspace{2.8cm}
\begin{abstract}
We construct the integral equations by taking the thermodynamic limit of
both the all-loop gauge Bethe ansatz equation and the string 
Bethe ansatz equation for the highest states in the 
$su(1|1)$ and $su(2)$ sectors of the
$\mathcal{N}=4$ super Yang-Mills theory. Using the Fourier transformation
we solve the integral equations iteratively to obtain the anomalous
dimensions of the highest states in the weak coupling expansion.
In the $su(1|1)$ sector we analytically study the strong coupling
limit of the anomalous dimension for the all-loop gauge Bethe ansatz
equation by means of the Laplace transformation.
\end{abstract}
\vspace{3cm}
\begin{flushleft}
March, 2007 
\end{flushleft}

\newpage
\section{Introduction}

The AdS/CFT correspondence \cite{MGW} has been deeply revealed by 
comparing the anomalous dimensions of certain single trace operators in
the planar limit of the $\mathcal{N}=4$ super Yang-Mills (SYM) theory
and the energies of certain string states in the type IIB string theory on
$AdS_5 \times S^5$ \cite{BMN,GKP,FT}.
In particular, integrability has made its appearance in both theories and
has shed light on the AdS/CFT correspondence.

The spectrum of anomalous dimension for a local composite operator
in the $\mathcal{N}=4$ 
SYM theory has been computed by the Bethe ansatz
\cite{MZ} for diagonalization of the dilatation operator
\cite{BKS,BZP} that is represented by a Hamiltonian of an integrable spin
chain with length $L$. Further, the asymptotic all-loop 
gauge Bethe ansatz (GBA) equations
for the integrable long-range spin chains have been proposed
in the $su(2), su(1|1)$ and $sl(2)$ sectors \cite{BDS,MS,BS}.

The integrability for the classical  $AdS_5 \times S^5$ string
sigma model has been investigated by verifying the equivalence between
the classical string Bethe equation for the string sigma model and
the Bethe equation for the spin chain \cite{KMM,VKZ}.
Combining the classical string Bethe ansatz and the asymptotic all-loop
GBA, a set of discrete Bethe ansatz equations for the quantum string
sigma model have been constructed \cite{AFS,NB,BS}, where the integrable
structure is assumed to be maintained at the quantum level and
the quantum string Bethe ansatz (SBA) equations are obtained by
modifying the GBA equations with the dressing factor. To fix the dressing
factor the SBA equation has been studied by comparing its prediction with
the quantum world-sheet correction to the spinning 
string solution \cite{SZZ}.
An all-order perturbative expression for the dressing factor at strong
coupling has been proposed \cite{BHL} such that it satisfies 
the crossing relation \cite{RJ} and matches with the known physical data
at strong coupling \cite{HL}. 

The spectrum of the highest state has been studied by analyzing the GBA
equation in the thermodynamic limit  $L \rightarrow \infty$  for the
$su(2)$ sector \cite{KZ} and the GBA and SBA equations for the $su(1|1)$
sector \cite{AT}. The flow of the spectrum from weak to strong coupling 
has been numerically derived by solving the GBA and SBA equations for the
$su(2)$ and $su(1|1)$ sectors in the large but finite $L$ \cite{BD}.
The strong coupling behavior of the su(2) spectrum has been investigated 
by using the Hubbard model which is regarded as the microscopic model
behind the integrable structure of the $\mathcal{N}=4$ SYM dilatation
operator \cite{RSS}. The highest states for the $su(2)$ and $su(1|1)$ 
sectors of the $AdS_5 \times S^5$ superstring have been studied 
analytically in the framework of the light-cone Bethe ansatz 
equations \cite{BAD}.

For the $sl(2)$ sector the large-spin anomalous dimension of twist-two 
operator has been computed by solving the GBA equation and  the SBA
equation in the thermodynamic limit by means of the Fourier transformation
\cite{ES}. In the former integral equation which we call the ES equation,
the anomalous dimension leads to the universal all-loop scaling function
$f(g)$ with the gauge coupling constant $g$ satisfying  the 
Kotikov-Lipatov transcendentality \cite{AKL}, whereas in the latter 
integral equation $f(g)$ is modified at the three-loop order as compared
to the ES equation and the transcendentality is not preserved.
In the GBA equation with a weak-coupling dressing factor 
\cite{BES} which is an 
analytic continuation of a crossing-symmetric strong-coupling dressing 
factor \cite{BHL}, which is called the BES equation, the universal
scaling function has been shown to be so modified at the four-loop order
as to obey the Kotikov-Lipatov transcendentality and be 
consistent with the planar multi-gluon amplitude of
$\mathcal{N}=4$ SYM theory at the four-loop order \cite{BCD}. 
The strong coupling behavior of $f(g)$ for the BES equation has been 
studied numerically by analyzing the equivalent set of linear algebraic 
equation to reproduce the asymptotic form predicted by the string 
theory \cite{BBK}. By truncating the strong coupling expansion of the 
matrices entering the linear algebraic equation, the strong coupling limit
of $f(g)$ has been extracted analytically \cite{AAB}.
In ref. \cite{KL} the ES and BES equations have been analyzed by using the
Laplace transformation where the analytic properties of the solutions 
at strong coupling are studied
and the strong coupling limit of $f(g)$ is estimated
analytically by deriving a singular solution for the integral equation.

Further for the $su(2)$ and $su(1|1)$ sectors the GBA equations with the
weak-coupling dressing factor have been analyzed and the anomalous 
dimensions of the highest states have been presented in the weak
coupling expansion \cite{RSZ}, where the anomalous dimensions of a 
state built from a field strength operator and 
a certain one-loop $so(6)$ singlet state also
have been computed. The physical origin of the full weak-coupling
dressing factor has been argued \cite{SS}. Without resorting to the 
Fourier transformation the strong
coupling solutions for the SBA equations in the rapidity  plane have
been analytically derived for the highest 
states in the $su(2)$ and $su(1|1)$
sectors and the strong coupling limit of the universal scaling function
$f(g)$ in the $sl(2)$ sector has been estimated from the BES equation
by deriving the leading density of Bethe roots in the rapidity
 plane \cite{KSV}.  On the other hand the Fourier-transform
of the SBA equation for the $sl(2)$ sector has been analyzed to
study the strong coupling behavior of $f(g)$ \cite{BDF}.

We will analyze the SBA equations for the highest states in the 
thermodynamic limit $L \rightarrow \infty$ for the $su(1|1)$ and
$su(2)$ sectors. By solving these equations through the Fourier 
transformation we will derive the anomalous dimensions of the highest 
states in the weak coupling expansion. Specially  the  weak coupling 
spectrum for  the $su(1|1)$ sector derived by computing the 
Fourier-transfomed density of Bethe roots 
will be compared with the result \cite{AT,BD}
which was produced by analyzing the SBA equation in the large but finite
$L$ and computing the Bethe momenta. Applying the Laplace transformation
prescription of ref. \cite{KL} to the GBA equation 
for the $su(1|1)$ sector we will construct a singular solution for the 
integral equation at strong coupling to compute the strong coupling limit
of the anomalous dimension analytically. 

\section{Weak coupling spectrum of the highest state in the
 $su(1|1)$ sector}

We consider the highest state in the $su(1|1)$ sector which corresponds to
the purely-fermionic operator $\mathrm{tr}(\psi^L)$ \cite{AT}, 
where $\psi$ is the
highest-weight component of the Weyl spinor from the vector multiplet.
The asymptotic all-loop GBA equation \cite{BS} for the highest state
is given by
\begin{equation}
\left(\frac{x_k^+}{x_k^-}\right)^L = \prod_{j\neq k}^L 
\frac{1- g^2/2x_k^+ x_j^-}{1- g^2/2x_k^- x_j^+},
\hspace{1cm} g^2 = \frac{\lambda}{8\pi^2},
\label{gba}\end{equation}
where $u_k \; (k=1, \cdots, L)$ are rapidities of elementary 
excitations and
\begin{equation}
x_k^{\pm} = x^{\pm}(u_k) = \frac{u_{\pm}}{2} \left( 1 +
\sqrt{ 1- \frac{2g^2}{u_{\pm}^2} }\right), 
\hspace{1cm} u_{\pm} = u_k \pm \frac{i}{2}.
\end{equation}
The all-loop ansatz (\ref{gba}) is a generalization of a three-loop Bethe
ansatz \cite{MS} and is obtained by deforming the spectral parameter $u_k$
into $x_k^{\pm}$ in such a way as $u_k \pm i/2 = x_k^{\pm} + 
g^2/2x_k^{\pm}$, where the deformation parameter is the Yang-Mills 
coupling constant $g$. The asymptotic all-loop energy $E(g)$ of the
highest state is 
\begin{equation}
E(g) = g^2 \sum_{k=1}^{L} \left( \frac{i}{x^+(u_k)} - 
 \frac{i}{x^-(u_k)} \right),
\label{egl}\end{equation}
which gives its dimension $\Delta = 3L/2 + E(g)$.
Taking the thermodynamic limit in the logarithm of
(\ref{gba}) and differentiating in
the rapidity $u$ we have an integral equation for the density
of Bethe roots $\rho(u)$ \cite{AT}
\begin{equation}
\frac{1}{i}\left( \frac{1}{\sqrt{u_+^2 - 2g^2}} - 
\frac{1}{\sqrt{u_-^2 - 2g^2}} \right) = -2\pi \rho(u) - \frac{i}{2} 
\int_{-\infty}^{\infty}dv \rho(v)\frac{\partial}{\partial u} \log \left(
\frac{1- g^2/2x^+(u)x^-(v)}{1- g^2/2x^-(u)x^+(v)} \right)^2
\label{ind}\end{equation}
where $u_{\pm}= u \pm i/2$  and  in the second term the density is 
integrated against the kernel
\begin{equation}
K_m(u,v) = i\frac{\partial}{\partial u} \log \left(
\frac{1- g^2/2x^+(u)x^-(v)}{1- g^2/2x^-(u)x^+(v)} \right)^2.
\label{kme}\end{equation}
In this continuum limit the energy 
shift $E(g)$ (\ref{egl}) is also expressed as an integral representation
\begin{equation}
\frac{E(g)}{L} = i g^2 \int_{-\infty}^{\infty} du \rho(u) \left(
  \frac{1}{x^+(u)} -  \frac{1}{x^-(u)} \right).
\label{ene}\end{equation}

Following the Fourier transformation procedure in ref. \cite{ES}, 
we solve the integral equation to obtain $E(g)$.
The Fourier transform of the density $\rho(u)$ is defined by
\begin{equation}
\hat{\rho}(t) = e^{-|t|/2}\int_{-\infty}^{\infty} du e^{-itu}\rho(u).
\label{den}\end{equation}
We are interested in the symmetric density $\rho(-u) = \rho(u)$ so that
$\hat{\rho}(t)$ is also symmetric $\hat{\rho}(-t) =  \hat{\rho}(t)$.
Therefore the kernel $K_m(u,v)$ in (\ref{ind}) can be symmetrized under
the exchange $v  \leftrightarrow -v$
\begin{equation}
i\partial_u \log \left( \frac{1- g^2/2x^+(u)x^-(u)}{1- g^2/2x^-(u)x^+(u)}
\right)^2 \rightarrow  \frac{i}{2}\partial_u \log \left( 
\frac{(1- g^2/2x^+(u)x^-(v)) (1+ g^2/2x^+(u)x^+(v))}
{(1- g^2/2x^-(u)x^+(v))(1+ g^2/2x^-(u)x^-(v))} \right)^2,   
\end{equation}
which is further described by \cite{ES}
\begin{equation}
g^2\int_{-\infty}^{\infty}dt e^{iut} \int_{-\infty}^{\infty}dt' e^{ivt'}
|t|e^{-(|t| + |t'|)/2}\hat{K}_m(\sg |t|, \sg|t'|),
\label{mke}\end{equation}
whose $\hat{K}_m$ is expressed in terms of the Bessel functions as
\begin{equation}
\hat{K}_m(x, x') = \frac{J_1(x)J_0(x') - J_0(x)J_1(x')}{x-x'}.
\end{equation}
We use the expression (\ref{mke}) to take the Fourier transformation
as $e^{-|t|/2}\int_{-\infty}^{\infty}du e^{-itu}\times$equation 
(\ref{ind}) and obtain
\begin{equation}
\hat{\rho}(t) = \et \left( J_0(\sg t) - g^2|t|\int_0^{\infty}dt'
\hat{K}_m(\sg|t|,\sg t') \hat{\rho}(t') \right).
\label{inf}\end{equation}

By solving this integral equation iteratively we derive the transformed
density $\hat{\rho}(t)$ expanded in even powers of $g$ as
\begin{eqnarray}
\hat{\rho}(t) = \et \biggl( 1 - \frac{g^2}{2}(t^2 + |t|)
+ \frac{g^4}{16} ( t^4 + 2(|t|^3 -t^2 +8|t|) ) \nonumber \\
- \frac{g^6}{288} ( t^6 + 3(|t|^5 -2t^4 + 26|t|^3 - 60t^2
+ 348|t|) )   \nonumber \\
+ \frac{g^8}{9216}( t^8 + 4(|t|^7 - 3t^6 + 54|t|^5 - 246t^4
+ 2520|t|^3 - 7200t^2 + 37296|t| )) + \cdots \biggr).
\label{dex}\end{eqnarray}
In deriving this solution we have used the following expansion 
\begin{eqnarray}
\hat{K}_m(\sg|t|, \sg t') = \frac{1}{2} \biggl( 1 - \frac{g^2}{4}( t^2
- |t|t' + t'^2 ) + \frac{g^4}{48}(t^4 - 2|t|^3t' + 4t^2t'^2 - 2|t|t'^3
+ t'^4 ) \nonumber \\
- \frac{g^6}{1152}( t^6 - 3|t|^5t' + 9t^4t'^2 - 9|t|^3t'^3 + 9t^2t'^4
- 3|t|t'^5 + t'^6 ) + \cdots \biggr).
\label{kmt}\end{eqnarray}
The energy shift $E(g)$ (\ref{ene}) can be expressed
in terms of the transformed density through (\ref{den}) as
\begin{equation}
\frac{E(g)}{L} = 4g^2\int_0^{\infty} dt \hat{\rho}(t)\frac{J_1(\sg t)}
{\sg t}.
\label{eni}\end{equation}
The substitution of the weak coupling solution (\ref{dex}) into 
(\ref{eni}) yields the anomalous dimension of the highest state
\begin{equation}
\frac{E(g)}{L} = 2g^2 - 4g^4 + \frac{29}{2}g^6 - \frac{259}{4}g^8
+ \frac{1307}{4}g^{10} + \cdots,
\label{egg}\end{equation}
which reproduces the result of \cite{AT,BD}.
In ref. \cite{AT} the most naive approximation to an exact expression
for the dimension was guessed in a square-root form as
\begin{eqnarray}
\frac{\Delta_{fit}}{L} &=& 1 + \frac{1}{2}\sqrt{1 + 
 \frac{\lambda}{\pi^2} } \nonumber \\
&=& \frac{3}{2} + \frac{\lambda}{4\pi^2} - \frac{\lambda^2}{16\pi^4}
+\frac{32\lambda^3}{1024\pi^6} - \frac{320\lambda^4}{16384\pi^8}
+ \cdots,
\label{exa}\end{eqnarray}
which is compared with the following weak coupling expansion 
in $\lambda$ associated with the anomalous dimension (\ref{egg})
\begin{equation}
\frac{\Delta}{L} = \frac{3}{2} + \frac{\lambda}{4\pi^2} - 
\frac{\lambda^2}{16\pi^4} + \frac{29\lambda^3}{1024\pi^6} - 
\frac{259\lambda^4}{16384\pi^8} + \cdots.
\end{equation}

Now we analyze the SBA equation for the highest state
\begin{equation}
\left(\frac{x_k^+}{x_k^-}\right)^L = \prod_{j\neq k}^L 
\frac{1- g^2/2x_k^+ x_j^-}{1- g^2/2x_k^- x_j^+}\sigma^2(x_k,x_j),
\label{sbe}\end{equation} 
whose string dressing factor $\sigma(x_k,x_j)$ is defined by
\begin{equation}
\sigma(x_k,x_j) = \left(\frac{1- g^2/2x_k^+ x_j^-}{1- g^2/2x_k^- x_j^+}
\right)^{-1}\left( \frac{(1- g^2/2x_k^+ x_j^-)(1- g^2/2x_k^- x_j^+)}
{(1- g^2/2x_k^+ x_j^+)(1- g^2/2x_k^- x_j^-)}\right)^{i(u_k -u_j)}.
\label{sdr}\end{equation}
In the thermodynamic limit the SBA equation (\ref{sbe}) becomes an
integral equation for the density $\rho(u)$
\begin{eqnarray}
\frac{1}{i}\left( \frac{1}{\sqrt{u_+^2 - 2g^2}} - 
\frac{1}{\sqrt{u_-^2 - 2g^2}} \right) &=& -2\pi \rho(u) - 
\frac{1}{2}\int_{-\infty}^{\infty}dv K_m(u,v)\rho(v) \nonumber \\
&-& \int_{-\infty}^{\infty}dv ( K_s(u,v) - K_m(u,v))\rho(v), 
\label{inu}\end{eqnarray}
where the main kernel $K_m(u,v)$ is given by (\ref{kme}) and 
\begin{equation}
K_s(u,v) = -\partial_u(u-v)\log \left( 
\frac{(1- g^2/2x^+(u)x^-(v)) (1- g^2/2x^-(u)x^+(v))}
{(1- g^2/2x^+(u)x^+(v))(1- g^2/2x^-(u)x^-(v))} \right)^2. 
\label{ksu}\end{equation}  
The last term of the r.h.s. of (\ref{inu}) specifed by $K_s - K_m$
appears as a contribution from
the dressing factor, which is compared with (\ref{ind}).

In the same way as (\ref{inf}) the Fourier transformation of (\ref{inu})
leads to
\begin{eqnarray}
-\et 2\pi J_0(\sg t) &=& -2\pi\hat{\rho}(t) + \pi g^2|t| 
\et\int_{-\infty}^{\infty}dt' \hat{K}_m(\sg|t|,\sg|t'|)
\hat{\rho}(t') \nonumber \\
&-& e^{-|t|/2}\int_{-\infty}^{\infty}du e^{-itu}\int_{-\infty}^{\infty}dv
K_s(u,v)\rho(v).
\label{iks}\end{eqnarray}
Using the $v \leftrightarrow -v$ symmetrized form of $K_s(u,v)$ we make
the third term on the r.h.s. of (\ref{iks}) rewritten by \cite{ES}
\begin{equation}
-2\pi g^2 |t|\et \int_{-\infty}^{\infty}dt' \left( 
\hat{K}_m(\sg|t|,\sg|t'|) + \sg\tilde{K}(\sg|t|,\sg|t'|) \right)
\hat{\rho}(t'),
\label{ksm}\end{equation}
where 
\begin{equation}
\tilde{K}(x,x') = \frac{x(J_2(x)J_0(x') -  J_0(x)J_2(x'))}
{x^2 - x'^2}.
\end{equation}
Thus we obtain an integral equation for the transformed density
\begin{eqnarray}
\hat{\rho}(t) &=& \et \biggl( J_0(\sg t)  - g^2|t|\int_0^{\infty}dt'
\hat{K}_m(\sg|t|,\sg t') \hat{\rho}(t') \nonumber \\
&-& 2g^2|t|\int_0^{\infty}dt'\sg\tilde{K}(\sg|t|,\sg t') 
\hat{\rho}(t')\biggr).
\label{til}\end{eqnarray}
Comparing (\ref{til}) with (\ref{inu}) and (\ref{inf})
 we see that the last term in 
(\ref{til}) specified by $\sg\tilde{K}$ is attributed to the dressing
factor so that  $\sg\tilde{K}$ is called a dressing kernel.

In order to solve (\ref{til}) by taking 
the weak coupling expansion we first split the transformed density 
$\hat{\rho}(t)$ into a main part $\hat{\rho}_0(t)$
and a correction part $\delta\hat{\rho}(t)$ as
$\hat{\rho}(t) = \hat{\rho}_0(t) + \delta\hat{\rho}(t)$, where 
$\hat{\rho}_0(t)$ satisfies the GBA equation (\ref{inf}).
Therefore we have the following integral equation for
$\delta\hat{\rho}(t)$
\begin{eqnarray}
\delta\hat{\rho}(t) &=& -2g^2|t| \et \biggl( \int_0^{\infty}
dt'\sg \tilde{K}(\sg|t|,\sg t')\hat{\rho}_0(t') \nonumber \\
&+& \frac{1}{2}\int_0^{\infty}
dt' \hat{K}_m(\sg|t|,\sg t')\delta\hat{\rho}_0(t') +
\int_0^{\infty}dt'\sg \tilde{K}(\sg|t|,\sg t')
\delta\hat{\rho}_0(t') \biggr),
\end{eqnarray}
where the first  term on the r.h.s. is regarded as an inhomogenous
one with $\hat{\rho}_0(t')$ already known as (\ref{dex}).
Using the expansion (\ref{kmt}) for $\hat{K}_m(x,x')$ and the
following weak coupling expansion for $\tilde{K}(x,x')$
with $x = \sg|t|, x' = \sg t'$
\begin{equation}
\tilde{K}(x,x') = \frac{\sg|t|}{8} \left( 1 - \frac{g^2}{6}(t^2 + t'^2)
+ \frac{g^4}{96}(t^4 + 3t^2t'^2 + t'^4) + \cdots \right)
\label{tkx}\end{equation}
we determine $\delta\hat{\rho}(t)$ iteratively
\begin{eqnarray}
\delta\hat{\rho}(t) &=& \et \biggl( - \frac{g^4}{2}t^2 + 
\frac{g^6}{12}( t^4 + 11t^2 + 6|t|) \nonumber \\
&-& \frac{g^8}{192}(t^6 + 30t^4 + 24|t|^3 + 384t^2 + 704|t|)
+ \cdots \biggr).
\end{eqnarray}
Combining them we obtain the anomalous dimension $E(g)/L = E_0(g)/L +
\delta E(g)/L$ where the main part $E_0(g)/L$ is given by (\ref{egg}) and
the correction part $\delta E(g)/L$ is evaluated as
\begin{equation}
\frac{\delta E(g)}{L} = -2g^6 + \frac{44}{3}g^8 
- \frac{268}{3}g^{10} + \cdots,
\label{deg}\end{equation}
whose expansion starts from the three-loop order.
The summation of (\ref{egg}) and (\ref{deg}) yields the dimension of the
highest state 
\begin{equation}
\frac{\Delta}{L} = \frac{3}{2} + 2g^2 - 4g^4 + \frac{25}{2}g^6
- \frac{601}{12}g^8 + \frac{2849}{12}g^{10} + \cdots,
\end{equation}
which recovers the result of \cite{AT,BD}. Thus
we have solved the SBA equation in the thermodynamic limit $L \rightarrow
\infty$ to derive the Fourier-transformed density iteratively, whereas
in \cite{AT,BD} the Bethe momenta of excitations in a finite fixed
$L$ were iteratively derived.

In \cite{BES} the universal scaling function $f(g)$ in the $sl(2)$
sector was obtained from the all-loop GBA equation with a weak-coupling
dressing factor and $f(g)$ was shown to satisfy the Kotikov-Lipatov
transcendentality. Since the dressing factor is universal for  the
three rank-one sectors, we use it for the $su(1|1)$ sector.
 In the $sl(2)$ sector, if we compare the
integral SBA equation for the transformed density in \cite{ES}
with the integral GBA equation accompanied with the weak-coupling dressing
factor in \cite{BES}, we note that the dressing kernel 
$\sg\tilde{K}(x,x')$ for the former case corresponds to the dressing 
kernel $2\hat{K}_c(x,x')$ for the latter case, where  $\hat{K}_c(x,x')$
is given by
\begin{eqnarray}
 \hat{K_c}(x,x') &=& 2g^2 \int_0^{\infty}dt"K_1(x,\sg t")
\frac{t"}{e^{t"}-1}K_0(\sg t",x'),  \nonumber \\
K_0(x,x')&=& \frac{xJ_1(x)J_0(x') -  x'J_0(x)J_1(x')}{x^2 - x'^2},
\nonumber \\
 K_1(x,x')&=& \frac{x'J_1(x)J_0(x') -  xJ_0(x)J_1(x')}
{x^2 - x'^2}. 
\end{eqnarray}
Therefore by replacing $\sg\tilde{K}(x,x')$ in (\ref{til}) with
$2\hat{K}_c(x,x')$ we obtain an integral equation for the transformed
density  in the $su(1|1)$ sector
\begin{eqnarray}
\hat{\rho}(t) &=& \et \biggl( J_0(\sg t) - g^2|t|\int_0^{\infty}dt'
\hat{K}_m(\sg|t|,\sg t')\hat{\rho}(t') \nonumber \\
&-& 2g^2|t|\int_0^{\infty}dt'2\hat{K}_c(\sg|t|,\sg t')\hat{\rho}(t') 
\biggr).
\label{rkc}\end{eqnarray}
Recently this integral equation has been presented and iteratively
solved in ref. \cite{RSZ}, where the energy modification owing to the
weak-coupling dressing factor starts from the four-loop order.

\section{Weak coupling spectrum of the highest state in the
$su(2)$ sector}

We turn to the highest state in the $su(2)$ sector which is described by 
the antiferromagnetic operator $\mathrm{tr}(Z^{L/2}\Phi^{L/2})+ \cdots$
where $Z$ and $\Phi$ are charged scalar fields in the $\mathcal{N}=4$
supermultiplet. The asymptotic all-loop GBA equation for the 
highest state is
\begin{equation}
\left(\frac{x_k^+}{x_k^-}\right)^L = \prod_{j\neq k}^{L/2} 
\frac{x_k^+ - x_j^-}{x_k^- - x_j^+}
\frac{1- g^2/2x_k^+ x_j^-}{1- g^2/2x_k^- x_j^+},
\end{equation}
whose thermodynamic limit leads to \cite{KZ}
\begin{equation}
\frac{1}{i}\left( \frac{1}{\sqrt{u_+^2 - 2g^2}} - 
\frac{1}{\sqrt{u_-^2 - 2g^2}} \right) = -2\pi \rho(u) - 
2\int_{-\infty}^{\infty}dv \frac{\rho(v)}{(u-v)^2 + 1}.
\label{urh}\end{equation}
The Fourier transformation solves the integral equation (\ref{urh}) to 
give an exact expression of the transformed density
\begin{equation}
\hat{\rho}(t) = \frac{J_0(\sg t)}{e^{|t|} + 1},
\label{exd}\end{equation}
which yields the dimension of the highest state $\Delta = L + E(g)$ in
a closed form
\begin{equation}
\frac{E(g)}{L} = 4g^2 \int_0^{\infty}\frac{dt}{\sg t} 
\frac{J_0(\sg t)J_1(\sg t)}{e^t + 1}.
\label{exe}\end{equation}
We use the following representation of the Riemann zeta function
\begin{equation}
\zeta (n + 1) = \frac{1}{(1-2^{-n})n!}\int_0^{\infty}dt
\frac{t^n}{e^t + 1}
\label{zet}\end{equation}
to expand (\ref{exe}) in $g^2$
\begin{eqnarray}
\frac{E(g)}{L} &=& 2\log 2g^2 - \frac{9}{4}\zeta(3)g^4 
+ \frac{75}{8}\zeta(5)g^6 \nonumber \\
&-& \frac{11025}{256}\zeta(7)g^8 + \frac{112455}{512}\zeta(9)g^{10}
 + \cdots,
\label{elg}\end{eqnarray}
whereas the closed expression (\ref{exe}) can yield $E(g)/L =
\sqrt{\lambda}/\pi^2$ in the strong coupling limit.

Let us consider the SBA equation for the highest state
\begin{equation}
\left(\frac{x_k^+}{x_k^-}\right)^L = \prod_{j\neq k}^{L/2} 
\frac{x_k^+ - x_j^-}{x_k^- - x_j^+}
\frac{1- g^2/2x_k^+ x_j^-}{1- g^2/2x_k^- x_j^+}
\sigma^2(x_k,x_j),
\label{sba}\end{equation}
where the string dressing factor $\sigma(x_k,x_j)$ is given by
(\ref{sdr}).
The thermodynamic limit of (\ref{sba}) yields an integral 
equation for the density 
\begin{eqnarray}
\frac{1}{i}\left( \frac{1}{\sqrt{u_+^2 - 2g^2}} - 
\frac{1}{\sqrt{u_-^2 - 2g^2}} \right) &=& -2\pi \rho(u) - 
2\int_{-\infty}^{\infty}dv \frac{\rho(v)}{(u-v)^2 + 1} \nonumber \\
&-& \int_{-\infty}^{\infty}dv(K_s(u,v) - K_m(u,v))\rho(v),
\label{sui}\end{eqnarray}
where the kernels $K_m(u,v)$ and $K_s(u,v)$ are given by (\ref{kme})
and (\ref{ksu})  respectively. 
The Fourier transformation of (\ref{sui}) through (\ref{ksm}) gives an
integral equation for the transformed density  
\begin{equation}
( 1 + \et )\hat{\rho}(t) = \et \left( J_0(\sg t) - 2g^2|t|
\int_0^{\infty}dt'\sg \tilde{K}(\sg|t|,\sg t') \hat{\rho}(t') \right),
\label{hrt}\end{equation}
whose last term is the same as the last one in (\ref{til}) for the
$su(1|1)$ sector. By using the expansion (\ref{tkx}) of $\tilde{K}(x,x')$
the transformed density
is iteratively solved as $\hat{\rho}(t)= \hat{\rho}_0(t) + 
\delta\hat{\rho}(t)$ where the main part $\hat{\rho}_0(t)$ is given by 
(\ref{exd}) and the correction part $\delta\hat{\rho}(t)$ has the 
following weak coupling expansion
\begin{eqnarray}
\delta\hat{\rho}(t) &=& \frac{1}{e^{|t|} + 1} \biggl( -\frac{g^4}{2}
\log2\; t^2 +  \frac{g^6}{12}(\log2\;t^4 +  6\zeta(3)t^2 ) \nonumber \\
&-& \frac{g^8}{384}( 2\log2\; t^6 + 33\zeta(3)t^4 + 675\zeta(5)t^2
- 144\log2\zeta(3)t^2 ) + \cdots \biggr).
\label{drh}\end{eqnarray}
The substitution of (\ref{exd}) and (\ref{drh}) into (\ref{eni}) leads to
a separation $E(g)/L = E_0(g)/L + \delta E(g)/L$ where the main part 
$E_0(g)/L$ takes the expression (\ref{elg}) 
and the correction part $\delta E(g)/L$ is estimated as
\begin{eqnarray}
\frac{\delta E(g)}{L} &=& -\frac{3}{2}\log2\zeta(3)g^6 + \left( 
\frac{75}{8}\log2\zeta(5) + \frac{3}{2}\zeta(3)^2 \right)g^8
\nonumber \\
&-& \left( \frac{6615}{128}\log2\zeta(7) + \frac{945}{64}\zeta(5)
\zeta(3) - \frac{9}{8}\log2\zeta(3)^2 \right)g^{10} + \cdots.
\label{loe}\end{eqnarray}
Thus it is noted that the weak coupling expansion of the energy correction
induced by the string dressing factor starts from the three-loop order 
in the same way as (\ref{deg}).

Now for $\sigma(x_k,x_j)$ in (\ref{sba}) we use the weak-coupling dressing
factor of the BES equation in ref. \cite{BES}. From the 
expression (\ref{hrt}) we replace the dressing kernel $\sg\tilde{K}(x,x')$
by the dressing kernel $2\hat{K}_c(x,x')$ to obtain
\begin{equation}
( 1 + \et )\hat{\rho}(t) = \et \left( J_0(\sg t) - 2g^2|t|
\int_0^{\infty}dt'2\hat{K}_c(\sg|t|,\sg t') \hat{\rho}(t') \right),
\end{equation}
whose last term is the same as the last one in (\ref{rkc}).
Recently this integral equation has been derived and solved in 
ref. \cite{RSZ}, where the energy correction also starts from the
four-loop order and a kind of transcendentality is observed if a degree
of transcendentality is assigned to both the ``bosonic" $\zeta$-function 
(\ref{zet}) and the ``fermionic" $\zeta_a$-function defined by 
$\zeta_a(n+1) = (1-2^{-n})\zeta(n+1)$. On the other hand, 
the summation of (\ref{elg}) and (\ref{loe}) expressed in terms of
$\zeta_a(1)= \log2$ leads to the following dimension of 
the highest state
\begin{eqnarray} 
\frac{\Delta}{L} &=& \frac{3}{2} + 2\zeta_a(1)g^2 - \frac{9}{4}\zeta(3)g^4
+ \left(\frac{75}{8}\zeta(5) - \frac{3}{2}\zeta_a(1)\zeta(3) \right)g^6
 \nonumber \\
&+& \left( -\frac{11025}{256}\zeta(7) + \frac{75}{8}\zeta_a(1)\zeta(5)
 + \frac{3}{2}\zeta(3)^2 \right)g^8 \\
&+& \left( \frac{112455}{512}\zeta(9) -\frac{6615}{128}
\zeta_a(1)\zeta(7) - \frac{945}{64}\zeta(3)\zeta(5) + 
\frac{9}{8}\zeta_a(1)\zeta(3)^2 \right)g^{10} + \cdots, \nonumber
\end{eqnarray}
which shows that the kind of transcendentality is not preserved for the
SBA equation.

\section{Strong coupling solution for the GBA equation in the 
$su(1|1)$ sector}

Here using the Laplace transformation prescription in ref. \cite{KL}
we analyze the strong coupling behavior of all-loop GBA equation for
the $su(1|1)$ sector. The eq. (\ref{inf}) can be written in terms of
$\hat{\rho}(t) = \ep f(x), t = \ep x, \ep=1/\sg$ as 
\begin{equation}
\ep f(x) = e^{-t}\left( J_0(x) - \frac{t}{2}\int_0^{\infty}
dx'\hat{K}_m(x,x')f(x') \right) 
\label{efx}\end{equation}
for $t > 0$. The energy shift (\ref{eni}) is extracted by
taking the following $t \rightarrow 0$ limit
\begin{equation}
\frac{E(g)}{L} = -4 \lim_{t \rightarrow 0}\frac{\ep f(x)e^t - J_0(x)}{t}.
\label{lim}\end{equation}
We use the expansion $\hat{K}_m(x,x')= 2\sum_{n=1}^{\infty}nJ_n(x)
J_n(x')/xx'$ and perform the Laplace transformation of (\ref{efx})
through $\phi(j) = \int_0^{\infty}dxe^{-xj}f(x)$  to have
\begin{eqnarray}
\ep \sqrt{j^2 +1} \phi(j-\ep) &=& 1 - \ep\int_{-i\infty}^{i\infty}
\frac{dj'}{2\pi i}\phi(j') \sum_{n=1}^{\infty}\left( 
\frac{-\sqrt{j'^2 + 1} + j'}{ \sqrt{j^2 + 1} +j} \right)^n \nonumber \\
&=& 1 - \ep\int_{-i\infty}^{i\infty}
\frac{dj'}{2\pi i}\phi(j') \frac{-\sqrt{j'^2 + 1} + j'}
{\sqrt{j^2 + 1} + j + \sqrt{j'^2 + 1} - j'},
\end{eqnarray}
whose integration contour can be enclosed around the cut which is
located to the right of it at the interval $-i < j' < i$.
The anti-symmetrization of the integral kernel to extract
the square-root singularity yields
\begin{eqnarray}
\ep \sqrt{j^2 +1} \phi(j-\ep) &=& 1 + \ep
\int_{-i}^i \frac{dj'}{2\pi i}\phi(j') \biggl( 
\frac{-\sqrt{j'^2 + 1} + j'}{\sqrt{j^2 + 1} + j + \sqrt{j'^2 + 1} - j'}
\nonumber \\
 &-& \frac{\sqrt{j'^2 + 1} + j'}
{\sqrt{j^2 + 1} + j - \sqrt{j'^2 + 1} - j'} \biggr). 
\end{eqnarray}
This integral equation is further expressed in terms of the
variable $z = \sqrt{j^2 + 1} + j$ and the new function 
$\chi(z) = \phi(j)$ as
\begin{equation}
\ep\frac{z^2 + 1}{2z} \chi(z_{\ep}) = 1 - \frac{\ep}{2}\int_{-i}^i
\frac{dz'}{2\pi i} \frac{z'^2 +1}{z'^2} \chi(z')
\left(\frac{z'}{z - z'} + \frac{1/z'}{z + 1/z'} \right),
\label{ekz}\end{equation}
where the integration over $z'$ is taken along a unit circle in the
anti-clock direction from $-i$ to $i$, and $z_{\ep}$ is defined by
\begin{equation}
z_{\ep} = \left( \Bigl( \frac{z^2 - 1}{2z} - \ep \Bigr)^2 + 1 
\right)^{1/2} + \frac{z^2 -1}{2z} - \ep.
\label{zez}\end{equation}
The transformation $z=z(j)$ provides the conformal mapping from two sheets
of the Riemann surface in the $j$-plane for $\phi(j)$ to one sheet in the
$z$-plane for $\chi(z)$. The eq. (\ref{ekz}) is rewritten by
\begin{equation}
\ep\frac{z^2 + 1}{2z} \chi(z_{\ep}) = 1 - \frac{\ep}{2}\int_{L}
\frac{dz'}{2\pi i} \frac{z'^2 +1}{z'} \frac{\chi(\tilde{z'})}{z- z'},
\label{ecl}\end{equation}
where the integration contour $L$ is given by a unit circle in the
anti-clock direction and $\tilde{z}$ is defined by 
$\tilde{z}_{Rez>0}=z, \; \tilde{z}_{Rez<0}=-z^{-1}$.
Here we assume that $\chi(\tilde{z'}) = \chi(z')$ in (\ref{ecl}),
that is, the symmetry of $\chi(z')$ under the substitution 
$z' \rightarrow -1/z'$ which means an analytic continuation of the
function $\phi(j')$ on the second sheet of the $j'$-plane with the 
substitution $\sqrt{j'^2 +1} \rightarrow - \sqrt{j'^2 +1}$.
Then we have 
\begin{equation}
\ep\frac{z^2 + 1}{2z} \chi(z_{\ep}) = 1 - \frac{\ep}{2}\int_{L}
\frac{dz'}{2\pi i} \frac{z'^2 +1}{z'} \frac{\chi(z')}{z- z'}.
\end{equation}

For the singular part of $\chi(z)$ inside the circle $|z|< 1$ we
obtain the following relation
\begin{equation}
\ep\frac{z^2 + 1}{2z}\chi(z_{\ep})_{sing} = 1 - 
\ep\frac{z^2 +1}{2z}\chi(z)_{sing}.
\label{eki}\end{equation}
In the strong coupling limit $\ep \rightarrow 0$ the eq. (\ref{zez})
becomes $z_{\ep} = z - 2z^2\ep/(1+z^2) + \cdots$ which makes
the relation (\ref{eki}) transformed into a first-order
differential equation
\begin{equation}
-\ep^2 \frac{\partial\chi(z)_{sing}}{\partial z} + 
\ep\frac{z^2 + 1}{z^2}\chi(z)_{sing} = \frac{1}{z}.
\label{fde}\end{equation}
The particular solution for this inhomogeneous differential equation
is obtained in the form of the expansion
\begin{equation}
\chi(z)_{sing}^{inhom} = \sum_{n=1}^{\infty}\frac{d_n}{z^n},
\end{equation}
where there is no regular term with $n=0$ and the coefficients are
specified by $d_1=1/\ep, \; d_2 = -1, \; d_3 = 2\ep - 1/\ep,
\cdots$. The homogeneous differential equation for (\ref{fde})
gives a solution
\begin{equation} 
\chi(z)^{hom} = C e^{\frac{1}{\ep}\left(z - \frac{1}{z}\right) }
= C\sum_{n=-\infty}^{\infty}z^nJ_n(2\ep^{-1}),
\end{equation}
where $C$ is an integral constant. The singular part of the 
solution $\chi(z)^{hom}$ is described by
\begin{equation}
\chi(z)_{sing}^{hom} = C\sum_{n=-\infty}^{-1}z^nJ_n(2\ep^{-1})
= C\sum_{n=1}^{\infty}\frac{(-1)^n}{z^n}J_n(2\ep^{-1}).
\end{equation}

Therefore the general solution for the first-order differential equation
(\ref{fde}) is represented by $\chi(z)_{sing}= \chi(z)_{sing}^{hom} +
\chi(z)_{sing}^{inhom}$ which is translated into
\begin{equation}
\phi(j) = \sum_{n=1}^{\infty}\frac{h_n}{j^n},
\end{equation}
where 
\begin{equation}
h_1 = \frac{1}{2}\left( \frac{1}{\ep} - CJ_1(2\ep^{-1}) \right),
\hspace{1cm} h_2 = \frac{1}{4}\left( -1 + CJ_2(2\ep^{-1}) \right), \cdots.
\end{equation}
The inverse Laplace transformtion leads back to
\begin{eqnarray}
f(x) &=& \int_{-i\infty}^{i\infty}\frac{dj}{2\pi i}
e^{xj}\phi(j) \nonumber \\
&=& \frac{1}{2}\left( \frac{1}{\ep} - CJ_1(2\ep^{-1}) \right) +
\frac{1}{4}\left( -1 + CJ_2(2\ep^{-1}) \right)\frac{t}{\ep}
+ \cdots.
\label{fxe}\end{eqnarray}
The integral constant $C$ is fixed by the requirement that
$E(g)/L$ in (\ref{lim}) should take a finite value in the limit 
$t \rightarrow 0$
\begin{equation}
C = -\frac{1}{\ep J_1(2\ep^{-1})}.
\end{equation}
The second term in (\ref{fxe}) yields the leading anomalous dimension
\begin{equation}
\frac{\Delta}{L}= \frac{1}{\ep}\frac{J_2(2\ep^{-1})}
{J_1(2\ep^{-1})},
\end{equation}
which is the contribution from the homogeneous differential equation and
is approximately expressed in the strong coupling region as
\begin{equation}
\lim_{g\rightarrow \infty} \frac{\Delta}{L} = \frac{\sqrt{\lambda}}{2\pi}
\tan\left( \frac{2}{\ep} - \frac{3}{4}\pi \right).
\label{gdl}\end{equation}

The resulting expression where the $g \rightarrow \infty$ limit is taken
after the $L \rightarrow \infty$ limit is compared with the estimation of
ref. \cite{BD}, where the Bethe momenta were computed at the fixed $L$
and strong $g$ region and then the strong anomalous dimension was 
evaluated numerically by choosing large $L$ 
\begin{equation}
\frac{\Delta}{L}= c_L \sqrt{\lambda}, \hspace{1cm}
c_L \rightarrow 0.1405 \; \mathrm{as} \; L \rightarrow \infty.
\end{equation}
Thus in the strong coupling limit we obtain the anomalous dimension
which rapidly oscillates around the estimation of ref. \cite{BD}.
This result for the GBA equation in the $su(1|1)$ sector resembles
the strong coupling behavior of the universal scaling function
$f(g)$ for the GBA equation in the $sl(2)$ sector presented
in ref. \cite{KL} where $f(g)$ oscillates around the value
predicted from the string theory.
Further we note that the factor $\sqrt{\lambda}/2\pi$ in (\ref{gdl})
coincides with the strong coupling limit of the conjectured square-root
formula (\ref{exa}).

\section{Conclusion}

We have investigated the SBA equations for the highest states in the
$su(1|1)$ and $su(2)$ sectors by applying the Fourier transformation
procedure in the rapidity  plane and using the expression of
the Fourier-transformed dressing kernel. We have computed the anomalous
dimensions of the highest states iteratively from the 
Fourier-transformed SBA equations and 
presented the alternative derivation of the anomalous dimension in the 
$su(1|1)$ sector which agrees with
the result of \cite{AT,BD}. The SBA equation in the 
thermodynamic limit $L \rightarrow \infty$ has been treated and the
Fourier-transformed density has been derived iteratively from the
integral equation, while in \cite{AT,BD} the SBA equation
in the large but finite $L$  has been analyzed and 
the Bethe momenta have been computed iteratively from the 
SBA equation in the momentum plane.

In the same manner as the SBA equation for the universal 
scaling function in the $sl(2)$ sector, 
we have demonstrated that for the SBA equation in the $su(2)$ sector
the contribution from the string dressing factor to the anomalous 
dimension starts from the three-loop order and there is 
a violation of the kind of transcendentality presented in ref. \cite{RSZ}
for the GBA equation with the weak-coupling dressing factor.

Following the Laplace transformation prescription we have analytically
studied the strong coupling behavior of the GBA equation for the
highest state in the $su(1|1)$ sector. The Laplace-transformed
GBA equation expressed as an integral equation has been changed into a
first-order differential equation in the strong coupling limit.
By constructing a singular solution for the differential equation
and taking the particular $t \rightarrow 0$ limit
we have extracted the strong coupling behavior of the anomalous
dimension and observed that it is mainly determined from 
the homogenous part of the differential equation.
 It has been shown that the analytically obtained
dimension oscillates around the value evaluated numerically from
the GBA equation in the momentum plane at large but finite
$L$ in ref. \cite{BD} and also the value estimated from the
square-root formula in ref. \cite{AT} conjectured by the
extrapolation of the weak-coupling expanded expression.


\begin{thebibliography}{99}
\bibitem{MGW} J.M. Maldacena, ``The large N limit of superconformal
field theories and supergravity,'' Adv. Theor. Math. Phys. \textbf{2}
(1998) 231 [hep-th/9711200]; S.S. Gubser, I.R. Klebanov and A.M. Polyakov,
``Gauge theory correlators from non-critical string theory,"
Phys. Lett. \textbf{B428} (1998) 105 [hep-th/9802109]; E. Witten, 
``Anti-de Sitter space and holography,"
Adv. Theor. Math. Phys. \textbf{2} (1998) 253 [hep-th/9802150].
\bibitem{BMN} D. Berenstein, J.M. Maldacena and H. Nastase, 
``Strings in flat space and pp waves from $\mathcal{N}$=4 super
Yang Mills," JHEP \textbf{04} (2002) 013 [hep-th/0202021].
\bibitem{GKP} S.S. Gubser, I.R. Klebanov and A.M. Polyakov,
``A semi-classical limit of the gauge/string correspondence,"
Nucl. Phys. \textbf{B636} (2002) 99 [hep-th/0204051].
\bibitem{FT} S. Frolov and A.A. Tseytlin, ``Semiclassical 
quantization of rotating superstring in $AdS_5\times S^5$," JHEP
\textbf{06} (2002) 007 [hep-th/0204226]; ``Multi-spin string solutions
in $AdS_5 \times S^5$," Nucl. Phys. \textbf{B668} (2003) 77 
[hep-th/0304255]; ``Rotating string 
solutions: AdS/CFT duality in non-supersymmetric sectors," Phys. Lett. 
\textbf{B570} (2003) 96 [hep-th/0306143];
A.A. Tseytlin, ``Spinning strings and AdS/CFT duality," hep-th/0311139.
\bibitem{MZ} J.A. Minahan and K. Zarembo, ``The Bethe-ansatz for 
$\mathcal{N} =4$ super Yang-Mills," JHEP \textbf{03} (2003) 013 
[hep-th/0212208].
\bibitem{BKS} N. Beisert, C. Kristjansen and M. Staudacher,
``The dilatation operator of conformal
$\mathcal{N} =4$ super Yang-Mills theory,"
Nucl. Phys. \textbf{B664} (2003) 131 [hep-th/0303060];
N. Beisert, ``The complete one-loop dilatation operator
of $\mathcal{N} =4$ super Yang-Mills theory," Nucl. Phys. \textbf{B676}
(2004) 3 [hep-th/0307015].
\bibitem{BZP} N. Beisert, ``The dilatation operator of 
$\mathcal{N} = 4$ super Yang-Mills theory and integrability," 
Phys. Rept. \textbf{405} (2005) 1 [hep-th/0407277];
K. Zarembo, ``Semiclassical Bethe ansatz and AdS/CFT," Comptes Rendus
Physique \textbf{5} (2004) 1081, Fortsch, Phys. \textbf{53}
(2005) 647 [hep-th/0411191];
J. Plefka, ``Spinning strings and integrable spin chains in the
AdS/CFT correspondence," Living Rev. Relativity \textbf{8}
(2005) 9 [hep-th/0507136].
\bibitem{BDS} N. Beisert, V. Dippel and M. Staudacher, ``A novel long
range spin chain and planar $\mathcal{N} =4$ super Yang-Mills,"
JHEP \textbf{07} (2004) 075 [hep-th/0405001].
\bibitem{MS} M. Staudacher, ``The factorized S-matrix of CFT/AdS,"
JHEP \textbf{05} (2005) 054 [hep-th/0412188].
\bibitem{BS} N. Beisert and M. Staudacher, ``Long-range PSU(2,2$|$4) Bethe
ansatze for gauge theory and strings," Nucl. Phys. \textbf{B727} (2005) 1
[hep-th/0504190].
\bibitem{KMM} V.A. Kazakov, A. Marshakov, J.A. Minahan and K. Zarembo, 
``Classical/quantum integrability in AdS/CFT," JHEP \textbf{05}
(2004) 024 [hep-th/0402207].
\bibitem{VKZ} V.A. Kazakov and K. Zarembo, 
``Classical/quantum integrability in non-compact sector of AdS/CFT," 
JHEP \textbf{10} (2004) 060 [hep-th/0410105];
N. Beisert, V.A. Kazakov and K. Sakai, ``Algebraic curve for the SO(6)
sector of AdS/CFT," Commun. Math. Phys. \textbf{263} (2006) 611
[hep-th/0410253];
S. Sch\"afer-Nameki, ``The algebraic curve of 1-loop planar 
$\mathcal{N} = 4$ SYM," Nucl. Phys. \textbf{714} (2005) 
3 [hep-th/0412254];
N. Beisert, V.A. Kazakov, K. Sakai and K. Zarembo,
``The algebraic curve of classical superstrings on $AdS_5 \times S^5$,"
Commun. Math. Phys. \textbf{263} (2006) 659 [hep-th/0502226].
\bibitem{AFS} G. Arutyunov, S. Frolov and M. Staudacher, ``Bethe
ansatz for quantum strings," JHEP \textbf{10} (2004)
016 [hep-th/0406256].
\bibitem{NB} N. Beisert, ``Spin chain for quantum strings," Fortschr. 
Phys. \textbf{53} (2005) 852 [hep-th/0409054].
\bibitem{SZZ} S. Sch\"afer-Nameki, M. Zamaklar and K. Zarembo, ``Quantum
corrections to spinning strings in $AdS_5 \times S^5$ and Bethe ansatz:
A comparative study," JHEP \textbf{09} (2005) 051 [hep-th/0507189]; 
N. Beisert and A.A. Tseytlin, ``On quantum corrections to spinning 
strings and Bethe equations," Phys. Lett. \textbf{B629} (2005) 102 
[hep-th/0509084];
S. Sch\"afer-Nameki and M. Zamaklar, ``Stringy sums and corrections
to the quantum string Bethe ansatz," JHEP \textbf{10} (2005) 044 
[hep-th/0509096]; 
S. Sch\"afer-Nameki, M. Zamaklar and K. Zarembo, ``How accurate is the
quantum string Bethe ansatz ?," JHEP \textbf{12} (2006)
020 [hep-th/0610250].
\bibitem{BHL} N. Beisert, R. Hernandez and E. Lopez, ``A 
crossing-symmetric phase for $AdS_5 \times S^5$ strings," JHEP \textbf{11}
(2006) 070 [hep-th/0609044].
\bibitem{RJ} R.A. Janik, ``The $AdS_5 \times S^5$ superstring worldsheet
S-matrix and crossing symmetry," Phys. Rev. \textbf{D73} (2006) 
086006 [hep-th/0603038].
\bibitem{HL} R. Hernandez and E. Lopez, ``Quantum corrections to the 
string Bethe ansatz," JHEP \textbf{07} (2006) 004 [hep-th/0603204];
G. Arutyunov and S. Frolov, ``On $AdS_5 \times S^5$ string S-matrix,"
Phys. Lett. \textbf{B639} (2006) 378 [hep-th/0604043];
L. Freyhult and C. Kristjansen, ``A universality test of the quantum
string Bethe ansatz," Phys. Lett. \textbf{B638} (2006) 258 
[hep-th/0604069].
\bibitem{KZ} K. Zarembo, ``Antiferromagnetic operators in $\mathcal{N}=4$
supersymmetric Yang-Mills theory," Phys. Lett. \textbf{B634} (2006) 552 
[hep-th/0512079].
\bibitem{AT} G. Arutyunov and A.A. Tseytlin, ``On highest-energy state in
the $su(1|1)$ sector of $\mathcal{N}=4$ super Yang-Mills theory,"
JHEP \textbf{05} (2006) 033 [hep-th/0603113].
\bibitem{BD} M. Beccaria and L. Del Debbio, ``Bethe ansatz solutions
for highest states in $\mathcal{N}=4$ SYM and AdS/CFT duality,"
JHEP \textbf{09} (2006) 025 [hep-th/0607236].
\bibitem{RSS} A. Rej, D. Serban and M. Staudacher ``Planar $\mathcal{N}=4$
gauge theory and the Hubbard model," JHEP \textbf{03} (2006) 018 
[hep-th/0512077]; 
J.A. Minahan, ``Strong coupling from the Hubbard model," J. Phys. 
\textbf{A39} (2006) 13083 [hep-th/0603175];
M. Beccaria and C. Ortix, ``Strong coupling anomalous dimensions of
$\mathcal{N}=4$ super Yang-Mills," JHEP \textbf{09}(2006) 016
[hep-th/0606138];
G. Feverati, D. Fioravanti, P. Grinza and M. Rossi, ``Hubbard's 
adventures in $\mathcal{N}=4$ SYM-land? Some non-perturbative 
considerations on finite length operators," J. Stat. Mech.
\textbf{02} (2007) P001 [hep-th/0611186].
\bibitem{BAD} M. Beccaria, G.F. De Angelis, L. Del Debbio and 
M. Picariello, ``Highest states in light-cone $AdS_5 \times S^5$
superstring," hep-th/0701167.
\bibitem{ES} B. Eden and M. Staudacher, ``Integrability and 
transcendentality," J. Stat. Mech. \textbf{11} (2006) P014
[hep-th/0603157].
\bibitem{AKL} A.V. Kotikov and L.N. Lipatov, ``DGLAP and BFKL equations
in the $\mathcal{N}=4$ supersymmetric gauge theories," Nucl. Phys.
\textbf{B661} (2003) 19; Erratum. ibid. \textbf{B685} (2004) 405
[hep-ph/0208220];
A.V. Kotikov, L.N. Lipatov and V.N. Velizhanin, ``Anomalous dimensions
of Wilson operators in $\mathcal{N}=4$ SYM theory," Phys. Lett.
\textbf{B557} (2003) 114 [hep-ph/0301021];
S. Moch, J.A.M. Vermaseren and A. Vogt, ``The three-loop 
splitting functions in QCD: The non-singlet case," Nucl. Phys.
\textbf{B688} (2004) 101 [hep-ph/0403192].
\bibitem{BES} N. Beisert, B. Eden and M. Staudacher, ``Transcendentality
and crossing," J. Stat. Mech. \textbf{01} (2007) P021 [hep-th/0610251].
\bibitem{BCD} Z. Bern, M. Czakon, L.J. Dixon, D.A. Kosower and 
V.A. Smirnov, ``The four-loop planar amplitude and cusp anomalous
dimension in maximally supersymmetric Yang-Mills theory," hep-th/0610248;
F. Cachazo, M. Spradlin and A. Volovich, ``Four-loop cusp anomalous
dimension from obtstructions," hep-th/0612309.
\bibitem{BBK} M.K. Benna, S. Benvenuti, I.R. Klebanov and A. Scardicchio,
``A test of the AdS/CFT correspondence using high-spin operators,"
hep-th/0611135.
\bibitem{AAB} L.F. Alday, G. Arutyunov, M.K. Benna, B. Eden and
I.R. Klebanov, ``On the strong coupling scaling dimension of high
spin operators," hep-th/0702028.
\bibitem{KL} A.V. Kotikov and L.N. Lipatov, ``On the highest
transcendentality in $\mathcal{N}=4$ SUSY," hep-th/0611204.
\bibitem{RSZ} A. Rej, M. Staudacher and S. Zieme, ``Nesting and
dressing," hep-th/0702151.
\bibitem{SS} K. Sakai and Y. Satoh, ``Origin of dressing phase
in $\mathcal{N}=4$ super Yang-Mills," hep-th/0703177. 
\bibitem{KSV} I. Kostov, D. Serban and D. Volin, ``Strong coupling
limit of Bethe ansatz equations," hep-th/0703031.
\bibitem{BDF} M. Beccaria, G.F. De Angelis and V. Forini, ``The
scaling function at strong coupling from the quantum string 
Bethe equations," hep-th/0703131.

\end{thebibliography}
\end{document}